\documentclass[a4paper]{jpconf}
\usepackage{graphicx}
\begin{document}

\def\sqrtsNN{\mbox{$\sqrt{s_\mathrm{NN}}$}}
\def\GeVc{\mbox{$\mathrm{GeV}/c$}}
\def\lt{\mbox{$<$}}
\def\gt{\mbox{$>$}}
\newcommand{ \be }{\begin{equation}}    
\newcommand{ \ee }{\end{equation}}    
\newcommand{ \bea }{\begin{eqnarray}}    
\newcommand{ \eea }{\end{eqnarray}}    {\Large {\normalsize }}
\newcommand{ \la }{\langle}    
\newcommand{ \ra }{\rangle}    

\title{Directed Flow of Identified Particles in Au + Au Collisions at $\sqrtsNN = 200$ GeV}

\author{Aihong Tang (for the STAR Collaboration)}

\address{Physics Department, Brookhaven National Laboratory, PO Box 5000, Upton, NY 11973}

\ead{aihong@bnl.gov}

\begin{abstract}
STAR's measurements of directed flow ($v_1$) for pions, kaons (charged and $K_s^0$), protons and antiprotons for Au + Au collisions at  $\sqrtsNN = 200$ GeV are presented. Negative $v_1(y)$ slope is observed for pions, antiprotons, protons and kaons.  The proton $v_1(y)$ slope at midrapidity is found extremely small. Sizable difference is seen between $v_1$ of protons and anti-protons in 5-30\% central collisions. Anti-flow can explain the negative slope, however, it has difficulties explaining the centrality dependence of the difference between the $v_1(y)$ slopes of protons and anti-protons. The $v_1$ excitation function is presented. Comparisons to model calculations (RQMD, uRQMD, AMPT and QGSM) are made, and it is found that none of the four models can successfully describe the data.
\end{abstract}

\section{Introduction}
In relativistic heavy ion collisions,  anisotropic flow describes the collective motion of particles with respect to the reaction plane, and it is conventionally characterized~\cite{methodPaper} by Fourier coefficients
\be
v_{n} = \la  \cos n (\phi - \psi) \ra
\label{vndef}
\ee
where $\phi$ denotes the azimuthal angle of an outgoing particle, $\psi$ is
the orientation of the reaction plane, and $n$ denotes the harmonic. So far three such coefficients are found finite at RHIC, namely, directed flow $v_1$, elliptic flow $v_2$ and 4$^{th}$ order harmonic flow $v_4$. This paper will focus on the directed flow, the first Fourier coefficient. Directed flow describes the sideward motion of fragments in ultra-relativistic nuclear collisions, and it carries early information from the collision~\cite{Sorge}. The shape of directed flow at midrapidity is of special interest because it might reveal a signature of a possible phase transition from normal nuclear matter to a Quark-Gluon Plasma (QGP)~\cite{antiFlow,thirdFlow,Stocker}. It is argued that directed flow, as a function of rapidity, may exhibit a flatness at midrapidity due to a strong, tilted expansion of the source. Such expansion gives rise to anti-flow~\cite{antiFlow} or a 3$^{rd}$ flow~\cite{thirdFlow} component. The anti-flow (3$^{rd}$ flow component) is perpendicular to the source surface, and is in the direction that is opposite to the direction of the bouncing-off motion of nucleons. If the tilted expansion is strong enough, it can even overcome the bouncing-off motion and results in a negative $v_1(y)$ slope at midrapidity, producing a wiggle-like structure.  In~\cite{Stocker}, it is emphasized that proton $v_1$ can serve as a probe for the first order phase transition, as calculation shows that proton $v_1(y)$ does not show negative slope for a hadronic Equation of State (EoS) without a QGP phase transition. Note that although the calculation is done with a first order phase transition at SPS energies, the direct cause of the negative slope is the strong, tilted expansion, which is also relevant at top RHIC energies. A wiggle structure is also seen in RQMD~\cite{wiggleRQMD}, and it is attributed to the baryon stopping and positive space-momentum correlation. In this picture, no phase transition is needed, and pions and nucleons  flow in opposite directions. To distinguish between baryon stopping and anti-flow associated with a phase transition, it is desirable to measure the $v_1(y)$ for identified particles and compare the signs of their slope at midrapidity. The centrality dependence of proton $v_1(y)$ may have implications for a possible first order phase transition. It is expected that in very peripheral collisions, protons flow in the same direction as spectators. In mid-central collisions, if there is a phase transition, the proton $v_1(y)$ slope at midrapidity may change sign and become negative. Eventually the slope diminishes in central collisions due to the symmetry of collisions. 

At low energies, it is shown by the E895 collaboration~\cite{e895KShort} that K$^0_s$ has a negative $v_1(y)$ slope around midapidity, while lambda and protons have positive slope~\cite{e895Lambda}. It is explained by a repulsive kaon-nucleon potential and attractive lambda-nucleon potential.  The NA49 collaboration~\cite{na49} has measured the $v_1$ for pions and protons, and negative $v_1(y)$ slope is observed by the standard event plane method. The three particle correlation method $v_1\{3\}$, which is believed to be less sensitive to non-flow effects, gives negative slope too, but with larger statistical error. At top RHIC energies, $v_1$ has been studied for charged particles by both the STAR and the PHOBOS collaborations~\cite{starV1V4,phobosV1,star62GeV,starV1PRL}. It is found that $v_1$ in the forward region follows the limiting fragmentation hypothesis well, and $v_1(\eta)$ depends only on the incident energy, but not on the size of the colliding system at a given centrality. The system size independence of $v_1$ can be explained by Hydrodynamic calculation with an titled initial condition~\cite{Bozek}. The systematic study of $v_1$ for identified particles at RHIC did not begin until recently because it is more challenging for two reasons, 1) $v_1$ for some identified particles (for example, protons) is much smaller than that of charged particles, thus is more difficult to measure , 2) it demands more statistics to measure $v_1$ for identified particles other than poins.

\section{Data Set}
In this analysis we used 62 million events for Au + Au collisions at $\sqrtsNN = 200$ GeV, all taken by STAR's minimum bias trigger during RHIC run 2007. Charged particles are identified by their energy loss inside STAR's Time Projection Chamber (TPC)~\cite{TPC}, and $K_s^0  (\rightarrow \pi^+ \pi^-)$ are reconstructed by their charged daughter tracks inside the TPC.  Track quality cuts are the same as used in~\cite{starFlowPRC}. The centrality definition of an event was based on the number of charged global tracks in the TPC with track quality cuts: $|\eta|<$0.5, a Distance of Closest Approach (DCA) to the global vertex less than 3 cm, and 10 or more fit points. Additional weight is assigned to each event in the analysis accounting for the bias in the vertex Z direction, due to the inclusion of hits of Silicon Vertex Tracker in tracking for year 2007 data production.  In addition, the transverse momentum $p_T$ for protons are required to be larger than 400 MeV$/c$, and DCA are required to be less than 1 cm, in order to avoid the beam-pipe background. The same cut is applied to anti-protons as well to enssure a fair comparison with protons. The high end of the $p_T$ cut is 1 GeV$/c$ where protons and pions have the same energy loss in the TPC and thus become indistinguishable. The event plane angle is determined from the sideward deflection of spectator neutrons measured by STAR's Shower Maximum Detector at Zero Degree Calorimeters (ZDCSMD), which are located close to beam rapidity and have the minimum contribution from non-flow. The description of measuring $v_1$ using the ZDCSMD event plane can be found in~\cite{star62GeV}.

\begin{figure}[htb]
\begin{center}
\includegraphics[width=28pc]{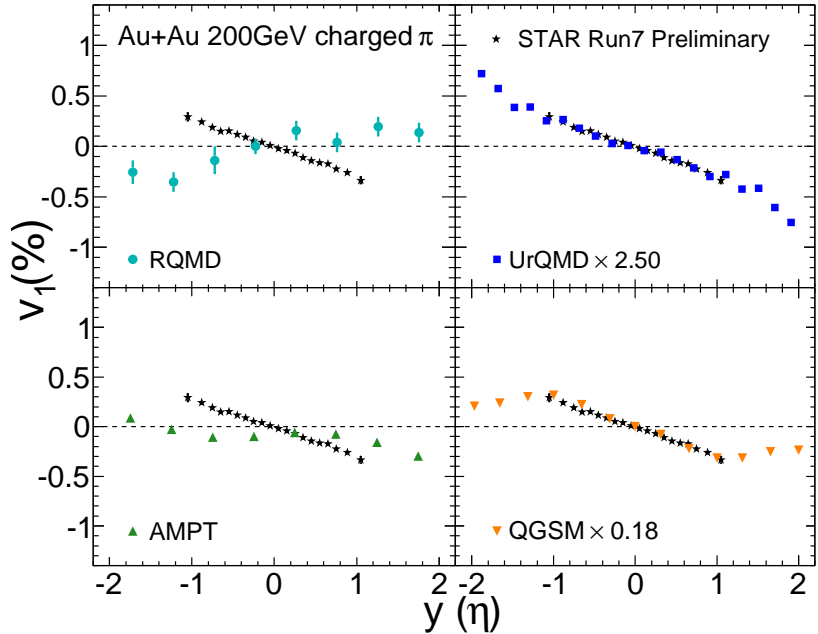}
\end{center}
\caption{\label{fig:pionV1}Pion $v_1$ as a function of rapidity (or $\eta$, for the QGSM model only) for 10-70\% Au + Au collisions at $\sqrtsNN = 200$ GeV. Also plotted are corresponding model calculations. The calculation from UrQMD and QGSM are scaled by a factor of 2.5 and 0.18, respectively. Errors are statistical only.}
\end{figure}

\section{Result}
Fig. \ref{fig:pionV1} shows the pion $v_1$ as a function of rapidity. The result is compared to four model calculations, namely, RQMD~\cite{wiggleRQMD}, UrQMD~\cite{UrQMD}, AMPT~\cite{AMPT} and QGSM~\cite{QGSM}.  Errors in this plot (and other data figures in this paper) are statistical only. Systematical uncertainties are studied by varying track quality cuts, and are found to be negligible. Calculations from RQMD and QGSM are scaled by a factor of 2.5 and 0.18, respectively. Following convention, the sign of spectator $v_1$ in the forward region is chosen to be positive, to which the measured sign of $v_1$ for particles of interest is only relative. The pion $v_1(y)$ result follows closely to the published charged particle result~\cite{star62GeV} and it shows a negative slope at midrapidity. Most models either predict the wrong sign of pion $v_1$ or the wrong magnitude by a signifiant factor~\footnote{However, in a recent paper~\cite{Bozek} posted after the $26^{th}$ Winter Workshop on Nuclear Dynamics, it is shown that the directed flow of charged particles can be explained by Hydrodynamic calculation with a tilted initial condition.}, which indicates that it is a challenging job to understand the dynamics happening at early stages. In this sense the measurement presented in this paper will offer strong constraints on models.
\begin{figure}[htb]
\begin{center}
\includegraphics[width=20pc]{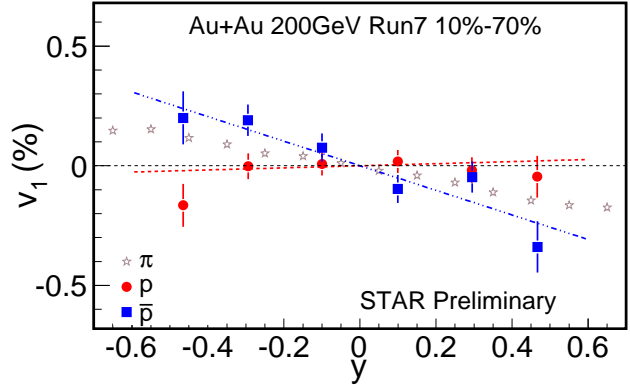}
\end{center}
\caption{\label{fig:PPbarPion}Proton (solid circles) and anti-proton (solid squares) $v_1$ as a function of rapidity for 10-70\% Au + Au collisions at $\sqrtsNN = 200$ GeV, pion $v_1(y)$ (empty stars) is shown as a reference. }
\end{figure}

To address the shadowing effect and the anti-flow, in Fig.\ref{fig:PPbarPion},  proton and anti-proton $v_1$ are plotted together with pion $v_1$. Pions are the dominant particles produced and they carry the flow of the bulk. It is not a surprise to see that anti-protons, also as produced particles, follow the flow direction of pions. They both have negative slope at midrapidity, which is consistent with the anti-flow picture. $v_1(y)$ of protons, however, exhibits a flatter shape than others. The mass difference between protons and pions cannot explain the flatness of proton $v_1(y)$ because the anti-protons have the same mass as that of protons, yet they flow close to pions. Indeed, the observed $v_1$ for protons is a convolution of directed flow of produced protons with that of transported protons, and the flatness of inclusive proton $v_1(y)$ around midrapidity can be explained by the negative flow of produced protons being compensated by the positive flow of protons transported from spectator rapidity, as a feature expected in the anti-flow picture. 
\begin{figure}[h]
\begin{center}
\includegraphics[width=20pc]{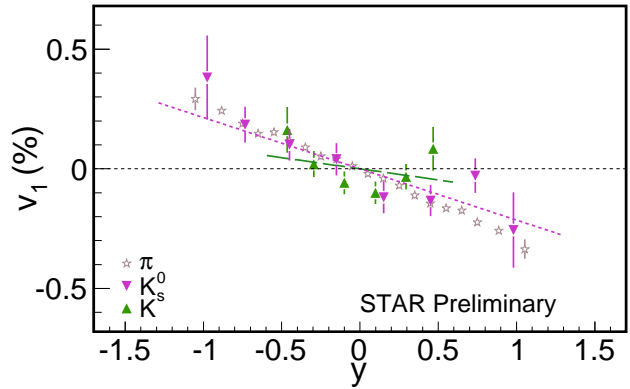}
\end{center}
\caption{\label{fig:kaonV1}Charged kaon (up triangles) and $K_s^0$ (down triangles) $v_1$ as a function of rapidity for 10-70\% Au + Au collisions at $\sqrtsNN = 200$ GeV, pion $v_1(y)$ (empty stars) is shown as a reference.}
\end{figure}
Fig. \ref{fig:kaonV1} shows $v_1$ of kaon and $K_s^0$ as a function of rapidity, for centrality 10-70\%. Pion $v_1$ is also plotted for comparison. Kaon $v_1$ is a unique measurement because the kaon/proton cross section is smaller than the pion/proton cross section, thus its directed flow is less likely to be "induced" by the shadowing effect. The plot shows that both kaon and $K_s^0$ have negative $v_1$ slope, which means that the observed $v_1$ is not likely to be caused by the shadowing effect. 

\begin{figure}[h]
\begin{center}
\includegraphics[width=20pc]{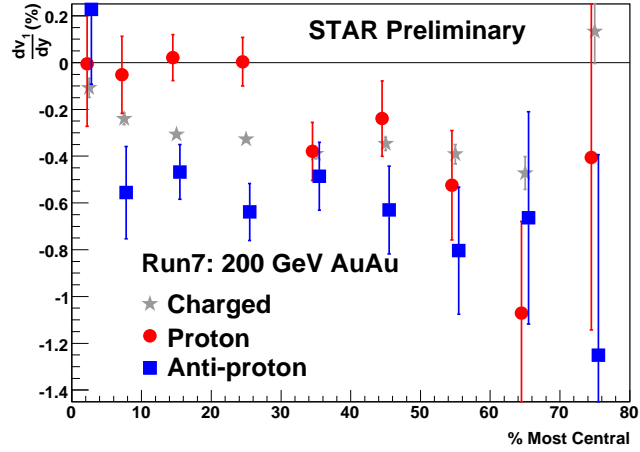}
\end{center}
\caption{\label{fig:v1Slope_cent}Charged (solid stars), proton (solid circles) and anti-proton (solid squares) $v_1(y)$ slope ($dv_1/dy$) at midrapidity as a function of centrality for Au + Au collisions at $\sqrtsNN = 200$ GeV.}
\end{figure}

On the other hand, anti-flow has difficulties in explaining the centrality dependence of $v_1$, as shown by Fig. \ref{fig:v1Slope_cent} , in which $v_1$ slope at midrapidity is plotted as a function of centrality for protons, anti-protons and charged particles.
If there is an anti-flow due to the strong, tilted expansion, one expects such an effect is larger in mid-central collisions than that in peripheral collisions. As a consequence, proton $v_1$ slope, which is expected to be positive in very peripheral collisions, will change its sign to negative in mid-central collisions and approach zero in central collisions. Due to the large statistical error, we cannot exclude the possibility that proton v1 can be positive in 70-80\% central collisions. However, in 30-80\% central collisions, proton $v_1$ slope is found mostly negative and the magnitude decreases with decreasing centrality. In more central (5-30\%) collisions, proton $v_1$ slope becomes extremely small, while anti-proton $v_1$ slope remains negative and continues to follow that of charged particles (mostly pions).  Anti-flow may cause a difference between $v_1$ of protons and anti-protons, and such difference is expected to be accompanied by strongly negative $v_1$.  In data, the large difference between proton and anti-proton $v_1$ is seen in 5-30\% centrality, while strongly negative $v_1$ is found for protons, anti-protons and charged particles in a different centrality (30-70\%). 
\begin{figure}[h]
\begin{center}
\includegraphics[width=28pc]{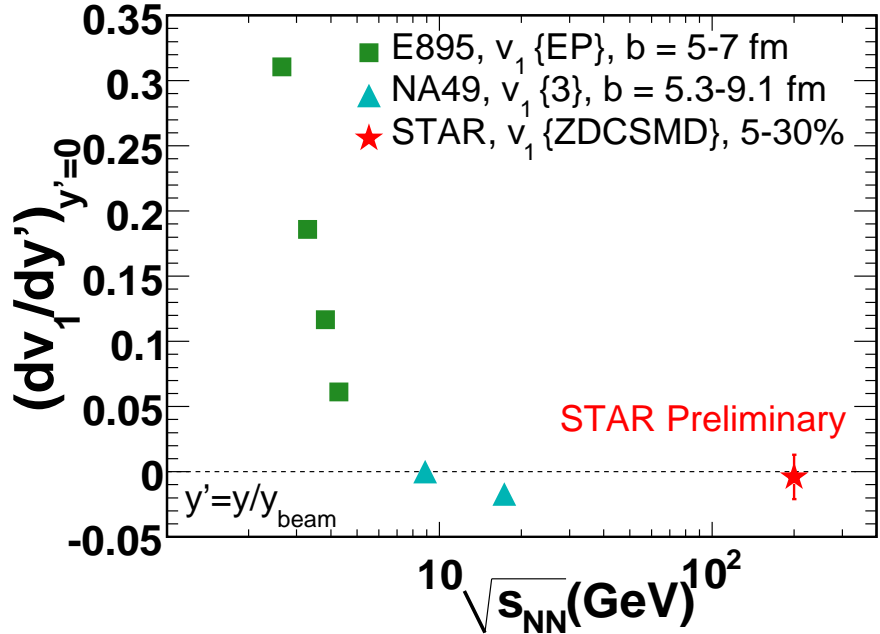}
\end{center}
\caption{\label{fig:v1ExcitationFcn}Proton $v_1(y^{'})$ slope ($dv_1/dy^{'}$) at midrapidity as a function of center of mass collision energy, where $y^{'}=y/y_{beam}$.}
\end{figure}

In Fig. \ref{fig:v1ExcitationFcn}, proton $v_1(y/y_{beam})$ slope at midrapidity is plotted as a function of collision energy. We see that proton $v_1$ slope decreases rapidly with increasing energy, reaching zero around $\sqrtsNN = 9$ GeV. It changes its sign to negative as shown by the data point at $\sqrtsNN = 17$ GeV, measured by NA49 experiment. With previous measurements, which include only one point above $\sqrtsNN = 9$ GeV, one cannot conclude if proton $v_1$ slope continues to decrease or stays close to zero when energy increases, the addition of the data point from RHIC indicates that proton $v_1$ slope remains close to zero at $\sqrtsNN = 200$ GeV. Judging over the broad energy range, the transition of the trend happens around $\sqrtsNN = 9$ GeV, making it a discontinuous point, and interestingly, it coincides with the energy vicinity where $\langle k^{+} \rangle/\langle \pi^{+} \rangle$ exhibits a kink~\cite{kink}. Hopefully the on-going Beam Energy Scan program will map out the region of interest with details.

\section{Summary}
For 10-70\% central Au + Au collisions at $\sqrtsNN = 200$ GeV, directed flow for pions, kaons (charged and $K_s^0$), and anti-protons are found to have negative slope at mid-rapidity. Protons, however, exhibit a flat shape around mid-rapidity. Sizable difference is seen between $v_1$ of protons and anti-protons in 5-30\% central collisions.  Anti-flow can explain the negative slope, however, it has difficulties explaining the centrality dependence of the difference between the $v_1(y)$ slopes of protons and anti-protons.

\end{document}